\documentclass[aps,preprint,nofootinbib,floatfix]{revtex4}
\usepackage{xcolor}

\pdfoutput=1
\usepackage{graphicx}
\usepackage[latin1]{inputenc}
\usepackage{amsmath,amssymb}
\usepackage{hyperref}
\usepackage{epstopdf}
\usepackage{slashed}
\usepackage{cancel}
\usepackage{soul}

\newcommand{\lsim}{\mathrel{\mathop{\kern 0pt \rlap
  {\raise.2ex\hbox{$<$}}}
  \lower.9ex\hbox{\kern-.190em $\sim$}}}
\newcommand{\gsim}{\mathrel{\mathop{\kern 0pt \rlap
  {\raise.2ex\hbox{$>$}}}
  \lower.9ex\hbox{\kern-.190em $\sim$}}}

 \def\be   {\begin{equation}}   \def\ee   {\end{equation}}
       \def\ea   {\end{array}}
 \def\bea  {\begin{eqnarray}}   \def\eea  {\end{eqnarray}}
 \def\bean {\begin{eqnarray*}}  \def\eean {\end{eqnarray*}}
 
 \def\Ga   {\Gamma}

     \def\De   {\Delta}

 \def\lee { \left( }
\def\rii { \right) }
\def\lan   {\langle}
\def\ran   {\rangle}

\def\tol {\leftrightarrow}

\begin{document}

{\small
\begin{flushright}
DO-TH 16/21 \\
\end{flushright} }

\title{Neutrino assisted GUT baryogenesis - revisited}
\author{
Wei-Chih Huang, Heinrich P\"as, Sinan Zei{\ss}ner
}

\affiliation{
\small{Fakult\"at f\"ur Physik, Technische Universit\"at Dortmund, 44221 Dortmund, Germany
}
}

\begin{abstract}
Many GUT models conserve the difference between the baryon and lepton number, $B-L$.
These models can create baryon and lepton asymmetries from heavy Higgs or gauge boson decays with $B+L \neq 0$ but with $B-L=0$. 
Since the sphaleron processes violate $B+L$, such GUT-generated asymmetries will finally be washed out completely, making GUT baryogenesis
scenarios incapable of reproducing the observed baryon asymmetry of the Universe.
In this work, we revisit the idea to revive GUT baryogenesis, proposed by Fukugita and Yanagida, where right-handed neutrinos 
erase the lepton asymmetry before the sphaleron processes can significantly wash out the original $B+L$ asymmetry, and in this way
one can prevent a total washout of the initial baryon asymmetry.
By solving the Boltzmann equations numerically for baryon and lepton asymmetries in a simplified 1+1 flavor scenario,
we can confirm the results of the original work.
We further generalize the analysis to a more realistic scenario of three active and two right-handed neutrinos to highlight flavor effects
of the right-handed neutrinos.  
Large regions in the parameter space of the Yukawa coupling and the right-handed neutrino mass
featuring successful baryogenesis are identified.
\end{abstract}

\maketitle

\section{Introduction}\label{sec:Introduction}
The unaccounted baryon asymmetry in the Universe is one of the reasons why the Standard Model~(SM) is imperfect and challenges physicists to come up with new ideas of asymmetry generation mechanisms.  
 Grand Unified Theories~(GUTs), on the other hand, elegantly unify all SM gauge couplings except for gravity and provide a seminal framework for baryogenesis model building. 
The simplest GUT based on the SU(5) gauge group proposed by Georgi and Glashow in 1974~\cite{Georgi:1974sy}, contains
 leptoquark gauge bosons which can mediate  baryon number violating processes, making the proton unstable, while preserving the difference between the baryon and the lepton number $B-L$.
 A baryon asymmetry can be generated from heavy gauge or Higgs boson decays \cite{Yoshimura:1978ex,Toussaint:1978br,Weinberg:1979bt,Barr:1979ye}, which fulfill all Sakharov conditions 
\cite{Sakharov:1967dj},
but is accompanied by a lepton asymmetry of equal amount.
 Thus the generated baryon and lepton asymmetries will be erased completely by the non-perturbative $B + L$ violating sphaleron
 processes~\cite{Klinkhamer:1984di, Arnold:1987mh,  Arnold:1987zg} which are effective when the temperature of the Universe falls below roughly $10^{12}$ GeV (while $B - L$ asymmetries are unaffected by the $B - L$ conserving sphalerons). The same problem also arises in other larger symmetry groups, such as $SO(10)$, which contains $U(1)_{B-L}$ as a subgroup.
In other words, as long as $U(1)_{B-L}$ is not broken when the baryon asymmetry is created, neither the baryon nor the lepton asymmetry will survive the sphaleron processes.

In principle there exist at least two remedies for GUT baryogenesis.
For certain matter representations under $SO(10)$ or larger groups
it has been demonstrated in Refs~\cite{Coughlan:1985hh,Babu:1992ia,Garbrecht:2005rr,Achiman:2007qz,Babu:2012vb,Babu:2012vc},
that a non-vanishing $B-L$ asymmetry can still be realized.
Alternatively, Fukugita and Yanagida \cite{Fukugita:2002hu} have proposed to  
involve right-handed neutrinos to revive GUT baryogenesis, where the right-handed neutrino $N$ can be a singlet under $SU(5)$ or be embedded into the $\bold{16}$ of $SO(10)$.
A Majorana mass of $N$, induced by a scalar vacuum expectation value~(VEV) or simply imposed by hand,
will explicitly violate the $B-L$ symmetry, that can either be a subgroup of the original GUT symmetry
or an accidental symmetry. 
The corresponding part of the Lagrangian reads
\begin{align}
 \mathcal{L}  \supset  y (\overline{\ell} \cdot H^*) N + \frac{M_N}{2} \overline{N^c} N + H.c. ,
\end{align}
where $\ell= \lee \nu_L \;  e_L \rii^T $ is the $SU(2)_L$ lepton doublet and  $H$ is the SM Higgs doublet.
Due to the Majorana mass term $M_N$ which violates the lepton number, 
the right-handed neutrinos can mediate $L$ washout processes such as
\begin{align}
\ell \ell \tol H^* H^*, ~~~  \ell H \tol \bar{\ell} H^*.
\label{processes1}
\end{align}
which destroy the initial lepton number asymmetry 
from GUT baryogenesis, while leaving the baryon number intact, resulting in a non-vanishing $B-L$ 
asymmetry which cannot be washed out by sphaleron effects.
In other words, the $N$-mediated washout processes generate a non-vanishing $B-L$ asymmetry from the
initial condition of a vanishing $B-L$ but finite $B +L$ asymmetry, making
a part of the baryon asymmetry immune against the action of the sphalerons. 
As a result, both the final $B$ and $L$ asymmetries can become nonzero.

\begin{figure}[!htb!]
\centering
\includegraphics[width=0.5\textwidth]{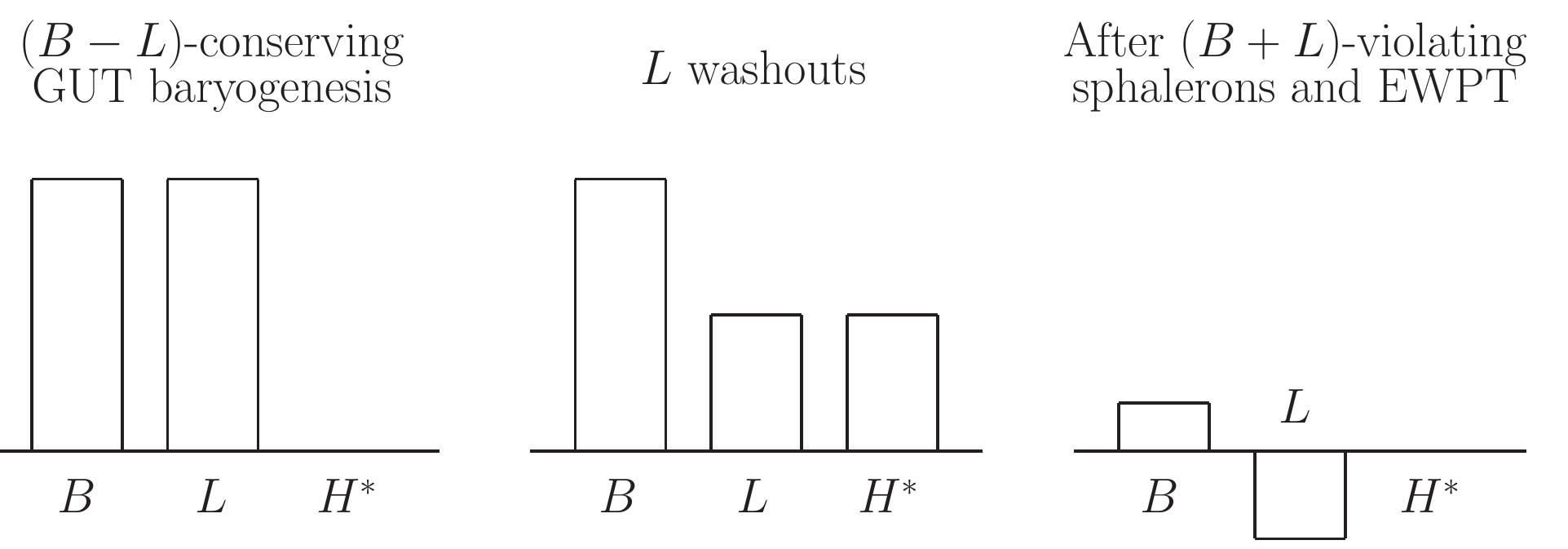} \\
\caption{Pictorial illustration of  asymmetry conversion. 
The left panel shows both $B$ and $L$ asymmetries created in equal amounts from GUT models which preserve $B-L$. 
In the center panel half of the $L$ asymmetry is converted into scalars $H^*$ via washout processes mediated by right-handed neutrinos.
The right panel shows how the sphalerons modify the $B$ and $L$ asymmetries while conserving $B-L$,
thus resulting in a nonzero $B$ asymmetry after the electroweak phase transition~(EWPT).
The $H$ asymmetry vanishes after the EWPT due to the Higgs VEV. (For explanation, see text)}
\label{fig:Asym_N}
\end{figure}

This entire process can create a non-zero baryon asymmetry after the electroweak phase transition, if
the $N$-induced washout processes are active before the onset of the sphaleron processes. It is important that
the sphalerons and $L$ washout processes do not coexist for too long since simultaneous $L$- and $(B+L)$-violating interactions
will erase both the $B$ and $L$ asymmetries.
Note that the SM Yukawa couplings, that couple right-handed to left-handed leptons via the Higgs boson,
are not in thermal equilibrium above a temperature of roughly $10^{12}$ GeV for $\tau$, $10^{9}$ GeV for $\mu$ and $10^{5}$ GeV for $e$.
As a consequence, the washout processes in Eq.~\eqref{processes1}
may only erase an initial asymmetry stored in the $SU(2)_L$ lepton doublets but not  in the right-handed charged leptons,
depending on the temperature. Throughout this work, for simplicity
we assume the $L$ asymmetry is stored in the lepton doublet only, i.e.,
a change on the number of the lepton doublet is equivalent to that of the lepton number: $\Delta \ell = \Delta L$.

For the $\Delta L=2$ washout processes, a change in the lepton number always comes with an equal amount of the $H^*$
number change, $\De L~(= - \De (B-L)) = \De H^*$, which can be easily seen
from Eq.~\eqref{processes1}. In addition, when these processes are in chemical equilibrium,
the chemical
potentials of $\ell$ and $H^*$ are equal.
Consequently, as shown in the center panel of Fig.~\ref{fig:Asym_N}
at most half of the initial $L$ asymmetry is converted into that of $H^*$
without considering impacts of the Yukawa couplings\footnote{For temperatures of interest,
the $t$~($T \lesssim 10^{16}$ GeV) or $b$~($T \lesssim 10^{12}$ GeV)
Yukawa interactions are effective during the $L$ washout.
The lepton asymmetry can be further shifted into a quark asymmetry
as discussed in Appendix~\ref{app:Boltz}, leading to larger final $(B-L)$ and $B$ asymmetries.}.
In other words, the maximal induced $B-L$ asymmetry by the $\Delta L=2$ interactions is one fourth of the initial $B+L$ asymmetry from GUT baryogenesis.
The SM Yukawa interactions later come into equilibrium and shuffle the asymmetry among quarks, leptons and Higgs bosons such that
the final  $B$ and $L$ asymmetries will be functions of the $B-L$ asymmetry, as indicated in the right panel of Fig.~\ref{fig:Asym_N}.
Note that as long as the $B-L$ asymmetry is non-vanishing,
the $B + L$ asymmetry will not be completely erased by the sphalerons~\cite{Khlebnikov:1988sr, Harvey:1990qw}.
That is due to the fact the sphalerons couple only to left-handed particles,
whereas the baryon and lepton numbers consist of both left-handed and right-handed
particles~(those are connected by the Yukawa couplings)\footnote{To be more precise, if the sphalerons and all the SM Yukawa couplings
are in equilibrium, all asymmetries can be expressed as functions of the lepton doublet asymmetry or its chemical potential, $\mu_\ell$. If $B-L \neq 0$, it implies $\mu_\ell \neq 0$
and thus $B+L \neq 0$.}.
After the electroweak phase transition~(EWPT), the $H^0$ asymmetry vanishes due to the Higgs VEV~\cite{Harvey:1990qw} 
$H^{\pm}$ and the imaginary part of the electrically neutral component are eaten by the $W^{\pm}$ and $Z$ bosons.

In this work, we revisit the idea of $N$-assisted GUT baryogenesis by numerically solving the Boltzmann equations including 
the lepton number violating ($\slashed{L}$) processes as well as the sphaleron effects, which allows a quantitative study of the relevant parameter space.
Special attention is paid to the investigation of the interplay between the washout and sphaleron processes, from which
one can infer the condition for obtaining the maximal final baryon and lepton asymmetries. 
We start with the case of one lepton generation and one right-handed neutrino, and then generalize this scenario to the realistic case of
three generations plus two right-handed neutrinos.

\section{Boltzmann equations for $L$ washout}\label{sec:boltz}

In this Section, we briefly discuss the Boltzmann equations used for obtaining the time evolution of the particle density in question.
More detailed discussions can be found in Refs.~\cite{Griest:1990kh,Edsjo:1997bg,Giudice:2003jh}.
The Boltzmann equation for a particle $\ell$
in the presence of the interaction $\ell a_1 \cdots a_n \leftrightarrow f_1 \cdots f_m$ is, 
\begin{align}
\label{eq:BoltzmannY}
	z H s \frac{Y_{\ell}}{dz} &=  \sum_{a_i,f_j} 
	\Big(\frac{n_{f_1} \cdots n_{f_m}}{n_{f_1}^{\rm eq} \cdots n_{f_m}^{\rm eq}}
      \gamma \left( f_1 \cdots f_m \leftrightarrow \ell a_1 \cdots a_n \right) \nonumber \\ & \qquad \qquad- \frac{n_{\ell} n_{a_1} \cdots n_{a_n}}{n_{\ell}^{\rm eq} n_{a_1}^{\rm eq} \cdots n_{a_n}^{\rm eq}}
      \gamma \left( \ell a_1 \cdots a_n \leftrightarrow f_1 \cdots f_m \right) \Big).
\end{align}
Here $Y~(\equiv n/s)$ denotes  
the particle number density normalized to the entropy density $s$ and $z = M_{N}/T$. 
The thermal equilibrium rate $\gamma$ is defined as
\begin{align}
	\gamma(\ell a_1 \cdots a_n \to f_1 \cdots f_m) 
	&=    \Big[ \int \frac{\mathrm{d}^3 p_{\ell}}{2 E_{\ell} (2\pi)^3} e^{-\frac{E_{\ell}}{T}} \Big] \prod\limits_{a_i} 
	\Big[ \int \frac{\mathrm{d}^3 p_{a_i}}{2 E_{a_i} (2\pi)^3} e^{-\frac{E_{a_i}}{T}} \Big] \nonumber\\ \times \prod\limits_{f_j} 
	\Big[ &\int \frac{\mathrm{d}^3 p_{f_j}}{2 E_{f_j} (2\pi)^3} \Big] 
        \times (2\pi)^4\delta^4 \Big( p_{\ell} + \sum_{i=1}^n p_{a_i} - \sum_{j=1}^m p_{f_j} \Big) |M|^2,
\label{eq:ga_def}         
\end{align}
with the squared amplitude, $|M|^2$, summing over initial and final spins. For a $2 \to 2$ process,
$\gamma$ can be further simplified if the corresponding
amplitude only depends on the square of the center-of-mass energy $s$ but not on the relative motion with respect to the thermal plasma:
\begin{align}
\gamma (a_1 a_2 \rightarrow f_1 f_2) = \frac{T}{64 \pi^4}\!\! \int \limits_{s_{min}}^\infty \!\! ds \sqrt{s} \ 
\hat{\sigma}(s) \ K_1 \left( \frac{\sqrt{s}}{T} \right) \, ,
\end{align}
with $s_{min} = \text{max}\lbrace (m_{a_1}+m_{a_2})^2, (m_{f_1}+m_{f_2})^2 \rbrace$ and the reduced cross-section $\hat{\sigma}(s)= 2 s \lambda(1,m_{a_1}^2/s,m_{a_2}^2/s) \sigma(s)$.

To simplify the analysis, we first focus on the $1+1$ scenario, one generation of SM leptons and one right-handed neutrino.
Moreover, we assume that the scale of GUT baryogenesis is below the right-handed neutrino mass to avoid complications from finite-temperature effects such as
$N \to H L$ being kinematically forbidden due to thermal masses when $T \gg M_{N}$~\cite{Giudice:2003jh}.
This scenario with low injection scales~($\lesssim M_N$) can be realized when the heavy particles responsible for baryogenesis are non-thermally produced as proposed
in Ref.~\cite{Kolb:1996jt}. Moreover, the $B-L$ asymmetry can also arise
even if the $B+L$ injection scale is higher than $M_N$ as long as the Yukawa coupling $y$ does not carry any $CP$ phase\footnote{If there exist
$CP$ phases, then the decay of $N$ is capable of yielding the lepton asymmetry, which is the typical mechanism for
leptogenesis~\cite{Fukugita:1986hr}, without resorting to GUT baryogenesis.},
and if lepton washout interactions and the sphalerons are not simultaneously effective for a long time.
In fact, for certain regions of the parameter space, a higher injection scale leads to a larger
$B-L$ due to a longer washout period.

To compute the $L$ washout, we include both $\Delta L=1$ and $\Delta L=2$ interactions. Following the same notation as used in Ref.~\cite{Giudice:2003jh},
the relevant $\De L =2$ washout processes are $\ell H \tol \bar{\ell} H^*$ and $\ell \ell  \tol H^* H^*$~(with thermal rates $\gamma_{Ns}$ and $\gamma_{Nt}$, respectively). The $\De L = 1$ washout processes include
$\ell H \tol N$ $\left(\gamma_D \right)$,
$\ell N \tol Q_3 \bar{U}_3$ $(\gamma_{Hs})$, $\ell \bar{Q}_3 \tol N \bar{U}_3$ and $\ell U_3 \tol N Q_3$ $(\gamma_{Ht})$, $\ell N \tol H^* A$ $(\gamma_{As})$, $\ell H \tol N A$ $(\gamma_{At_1})$
and $\ell A \tol N H^*$ $(\gamma_{At_2})$,
where $Q_3$ refers to the third generation quark doublet while $U_3$~($D_3$) denotes the right-handed top quark~(bottom quark).
Again, to evade complications from finite-temperature effects we focus on scenarios with regions of interest
$T \lesssim M_N$. Therefore, the $\Delta L=1$ interactions that involve an external right-handed neutrino $N$
are always Boltzmann suppressed compared to the $\Delta L=2$ processes.

As explained in Ref.~\cite{Giudice:2003jh}, for $\De L =2$ processes one has to subtract contributions from on-shell right-handed neutrinos $N$
to avoid double counting,  if the contributions of their decays and inverse decays
are already taken into account, i.e., the processes $\ell H \tol N \tol \bar{\ell} H^* $ have been included by successive decays.
Alternatively, one can simply consider $\Delta L=2$ interactions with on-shell right-handed neutrinos without including the (inverse) decays as they are already incorporated in the the unsubtracted rate. In this work, we adopt the second method.

Assuming no $CP$ violation sources in the $N$ decay as mentioned above,
the Boltzmann equation for the lepton asymmetry with  $ Y_{L} \equiv \left( Y_\ell - Y_{\bar{\ell}} \right) \ll Y_\ell$
reads 
\begin{align}
zHs \frac{d Y_{L}}{dz}  = &- 2 \left( 2 \gamma_{Ns} + 4 \gamma_{Nt} + \gamma_{Hs} \frac{Y_N}{Y_N^{eq}} + 2 \gamma_{Ht} +\gamma_{As} \frac{Y_N}{Y_N^{eq}} + \gamma_{At_1} + \gamma_{At_2} \right) \frac{Y_{L}}{Y_L^{eq}} \nonumber \\
 &+ 2 \bigg[ \bigg(\frac{Y_{H}^{eq}}{Y_{U_3}^{eq}} b_{U_3} -\frac{Y_{H}^{eq}}{Y_{Q_3}^{eq}} b_{Q_3}\bigg) \bigg( \gamma_{Hs} + \bigg( 1 + \frac{Y_N}{Y_N^{eq}} \bigg) \gamma_{Ht} \bigg) \nonumber \\
 &+2 b_H \gamma_{Ns} + 4 b_H \gamma_{Nt} +b_H \gamma_{As}+  b_H\gamma_{At_1} +  b_H\gamma_{At_2}  \frac{Y_N}{Y_N^{eq}}  \bigg] \frac{Y_{H^\prime}}{Y_{H}^{eq}} \; , \label{eq:dYL/dz}
\end{align}
with the Hubble parameter 
\begin{align}
H \simeq1.66 g^{1/2}_*  \, T^2/M_{pl} \, .
\end{align}
Similar to the definition of $Y_L$, $Y_i$~($i=Q_3,  U_3,  D_3,  H $) are defined as the
asymmetry densities\footnote{The symbol ``$\Delta$'' is reserved for the change of the particle number due to $L$ washouts, such as
$\Delta L$ and $\Delta H$.}, i.e., $Y_i \equiv (Y_{\text{particle $i$} }- Y_{\text{anti-particle $\bar{i}$}} ) $  whereas
$Y_i^{eq}$ always refer to the equilibrium density: $Y_i^{eq} \equiv Y_{\text{particle $i$} }^{eq}=Y_{\text{anti-particle $i$} }^{eq}$.
Here $M_{Pl}$ is the Planck mass and $g_*$ is the number of relativistic degrees of freedom~(106.75 for the SM). 
$Y_{H^{\prime}}$ is the $L$ asymmetry change due to washouts\footnote{In the absence of the Yukawa couplings,
$Y_{H^\prime=}Y_H$ since $\Delta L = \Delta H^* =- \Delta H$ as mentioned above.}, i.e., $Y_{H^{\prime}} (z) \equiv
Y^{\text{initial}}_{L} - Y_{L}(z)$ which implies,
\begin{align}
 \frac{d Y_{H^\prime}}{dz}  = -  \frac{d Y_{L}}{dz}  \; .
 \label{eq:dYH/dz}
\end{align}
As we shall see in Appendix~\ref{app:Boltz}, the impact of the heavy quark Yukawa couplings on the washout processes can be characterized by the factors $b_{(H,Q_3,U_3)}$:
\begin{align}
b_{(H,~Q_3,~U_3)} = \left\lbrace \begin{matrix}
\frac{1}{2},-\frac{1}{2},~\frac{1}{2} &\;\;  \;\; 10^{12} \lesssim T  \lesssim 10^{16}\text{ GeV}\\
\frac{1}{4},~ 0,~\frac{3}{8}  &\;\; \;\; T \lesssim 10^{12}\text{ GeV}
\end{matrix} \right. \;\; .
\end{align}
Note that we do not consider the $\tau$ Yukawa coupling, which is in equilibrium for $T \lesssim 10^{12}$ GeV. This is justified as the contribution is relatively small,
compared to those of $t$ and $b$ Yukawa couplings in light of the color factor.  

%

\begin{figure}[!htb!]
\centering
\includegraphics[width=0.45\textwidth]{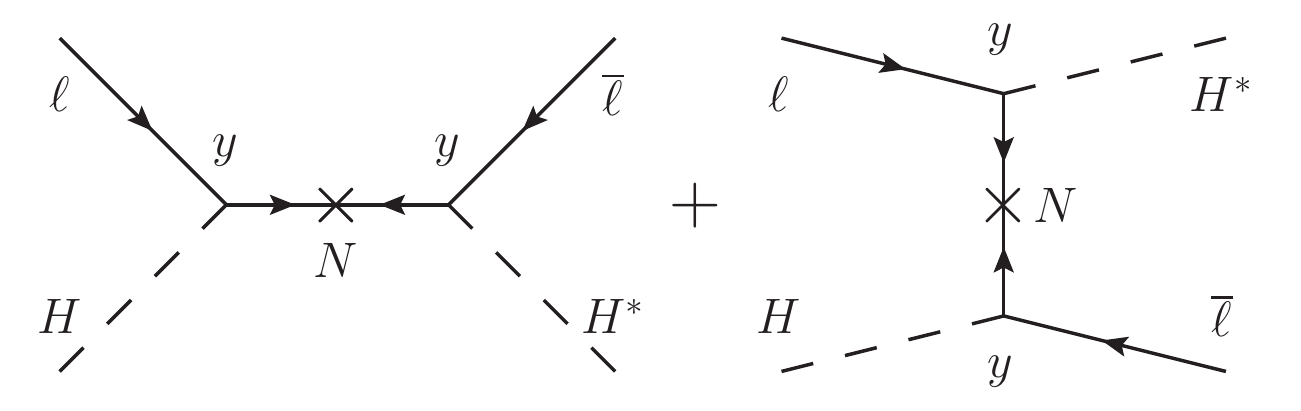}
\includegraphics[width=0.45\textwidth]{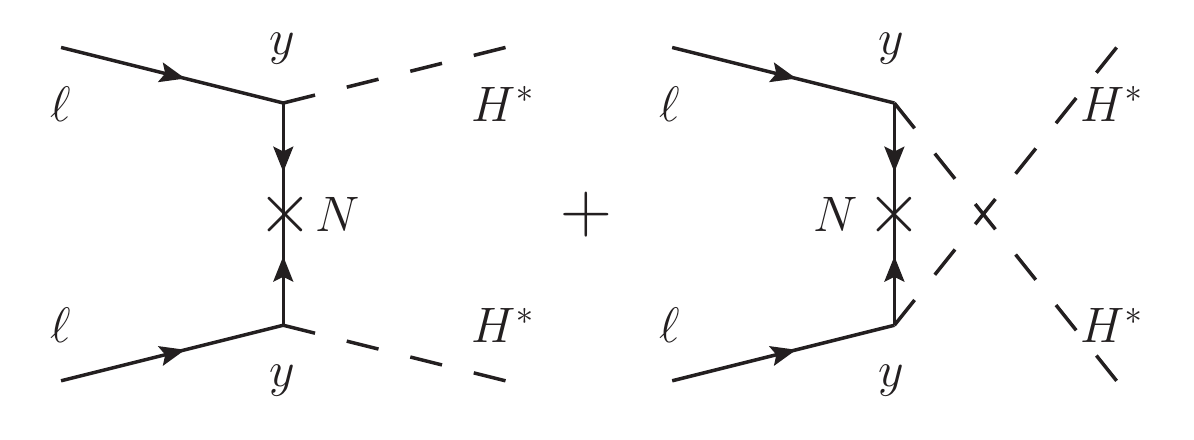}
\caption{ $N$-mediated lepton number violating processes.}
\label{fig:N_washout}
\end{figure}

Finally, the Boltzmann equation for the number density of $N$ is given by:
\begin{align}
zHs \frac{d Y_{N}}{dz}  =  &- \left( \gamma_D + 4\gamma_{Hs} + 8\gamma_{Ht} + 4\gamma_{As} + 4\gamma_{At_1} + 4\gamma_{At_2} \right) \left(\frac{Y_{N}}{Y_N^{eq}}-1\right) \; . \label{eq:dYN/dz}
\end{align}
As all reduced cross-sections $\hat{\sigma}$s for the $\Delta L=1$ interactions are given in Ref.~\cite{Giudice:2003jh},
we here only provide the reduced cross-sections $\hat{\sigma}_{Ns}$ and $\hat{\sigma}_{Nt}$ for the $\Delta L=2$ processes,
$\ell H \tol \bar{\ell} H^*$ and $\ell \ell \tol H H^*$ respectively, as displayed in Fig.~\ref{fig:N_washout}. 

The reduced cross section of $\ell H \tol \bar{\ell} H^*$ with the center-of-mass energy squared $s$ is:
\begin{align}
&\hat{\sigma}_{Ns} = M_N^2 2 u \lambda \left[ 1,\frac{a_\ell}{u},\frac{a_H}{u} \right] \sigma_{Ns}(u)= |y|^4 \frac{\lambda \left[ 1,\frac{a_\ell}{u},\frac{a_H}{u} \right]}{32 \pi u} \int \limits_{-1}^1 dx \big( (u+a_\ell-a_H)^2\nonumber \\&\,\,\,\,+(a_H(a_H-2a_\ell-2u)+ (u-a_\ell)^2)x\big) \bigg[ \frac{2}{a_\Gamma + (u-1)^2} + \left( \frac{8u^2}{\alpha^2}+  \frac{4u(u-1)}{(a_\Gamma + (u-1)^2) \alpha}\right) \bigg] \;\; ,
\end{align}
where
\begin{align}
\alpha= \big(u^2 + (a_H - a_\ell)^2 \big) x + 2 u \big(-1 + (a_H + a_\ell) (1-x)\big)-\sqrt{(u+a_H - a_\ell)^2(u+a_\ell - a_H)^2} \;\; ,
\end{align}
$ u= s /M^2_N$, $a_i = m_i^2/M_N^2$ and $x=\cos\theta$. Here, $\theta$ is the angle between the incoming and the outgoing lepton.
For $\ell \ell \tol H^* H^*$, the reduced cross section reads:
\begin{align}
\hat{\sigma}_{Nt} = 2 s \lambda \left[ 1,\frac{a_\ell}{u},\frac{a_\ell}{u} \right] \sigma_{Nt}= |y|^4 \frac{\lambda \left[ 1,\frac{a_\ell}{u},\frac{a_\ell}{u} \right]}{64 \pi} \int \limits_{-1}^1 dx (u-2a_L) \bigg[\frac{2}{\beta^2} + \left( \frac{1}{\beta}+ \frac{1}{\beta-(u-4a_H)x}\right)^2 \bigg] \;\; ,
\end{align}
with
\begin{align}
\beta = a_H+a_\ell-1+\frac{1}{2}\big((u-4a_H)x-\sqrt{u(u+4a_\ell-4a_H)} \big) \;\; .
\end{align}

\section{Sphalerons processes}\label{sec:spha_L}
We are now in the position to include the sphaleron processes into the Boltzmann equations
to study the interplay between the $\slashed{L}$ and $(B+L)$-violating (denoted by $\cancel{B+L}$) interactions.
The sphaleron effects can be expressed as~\cite{Moore:2000ara,D'Onofrio:2014kta} 
\begin{align}
\frac{\dot{Y}_{B}}{Y_{B} + Y_{L}}=\frac{\dot{Y}_{L}}{Y_{B} + Y_{L}}
= \frac{39}{8} \lee 18 \alpha_W^5 T \rii \; ,
\end{align}
where $Y_B \equiv Y_{\text{baryon}} - Y_{\text{anti-baryon}}$, $\dot{Y}=d Y/dt$ and $dt=0.6 \, g_*^{-1/2} \lee M_{Pl}/M^2_N \rii z dz$.
Clearly the sphalerons erase the $B$ and $L$ asymmetries at the same rate so that the $B-L$
asymmetry remains constant. Thus the sphalerons alter $Y_{B+L}$ but not $Y_{B-L}$ in the Boltzmann
equations:
\begin{align}
zHs \frac{d Y_{B- L}}{dz} = & 2 \left( 2 \gamma_{Ns} + 4 \gamma_{Nt} + \gamma_{Hs} \frac{Y_N}{Y_N^{eq}} + 2 \gamma_{Ht} +\gamma_{As} \frac{Y_N}{Y_N^{eq}} + \gamma_{At_1} + \gamma_{At_2} \right) \frac{Y_{B+L}-Y_{B-L}}{2Y_L^{eq}} \nonumber \\
 &- 2 \bigg[ \bigg(\frac{Y_{H}^{eq}}{Y_{U_3}^{eq}} b_{U_3} -\frac{Y_{H}^{eq}}{Y_{Q_3}^{eq}} b_{Q_3}\bigg) \bigg( \gamma_{Hs} + \bigg( 1 + \frac{Y_N}{Y_N^{eq}} \bigg) \gamma_{Ht} \bigg) \nonumber \\
 &+2 b_H \gamma_{Ns} + 4 b_H \gamma_{Nt} +b_H \gamma_{As}+ b_H\gamma_{At_1} + b_H\gamma_{At_2}  \frac{Y_N}{Y_N^{eq}}  \bigg] \frac{Y_{H'}}{Y_{H}^{eq}} \ , \label{Y_B-L}
 \end{align}
 \begin{align}
zHs \frac{d Y_{B+L}}{dz} = &- 2 \left( 2 \gamma_{Ns} + 4 \gamma_{Nt} + \gamma_{Hs} \frac{Y_N}{Y_N^{eq}} + 2 \gamma_{Ht} +\gamma_{As} \frac{Y_N}{Y_N^{eq}} + \gamma_{At_1} + \gamma_{At_2} \right) \frac{Y_{B+L}-Y_{B-L}}{2Y_L^{eq}} \nonumber \\
 &+ 2 \bigg[ \bigg(\frac{Y_{H}^{eq}}{Y_{U_3}^{eq}} b_{U_3} -\frac{Y_{H}^{eq}}{Y_{Q_3}^{eq}} b_{Q_3}\bigg) \bigg( \gamma_{Hs} + \bigg( 1 + \frac{Y_N}{Y_N^{eq}} \bigg) \gamma_{Ht} \bigg) +2 b_H \gamma_{Ns} + 4 b_H \gamma_{Nt}  \nonumber \\
 &+b_H \gamma_{As}+ b_H\gamma_{At_1} + b_H\gamma_{At_2}  \frac{Y_N}{Y_N^{eq}}  \bigg] \frac{Y_{H'}}{Y_{H}^{eq}} 
 + \frac{351}{2} \alpha_W^5 \frac{M_N s}{z} Y_{B+L} \ , \label{Y_B+L}\\
 \frac{d Y_{H'}}{dz} = &  \frac{d Y_{B-L}}{dz} \ \label{Y_H'} .
\end{align}
After all SM Yukawa interactions reach thermal equilibrium, the final baryon and lepton asymmetries are directly related to the $B-L$ asymmetry
created by the washout processes:
\begin{align}
Y^{\text{final}}_{B}= c_s Y^{\text{final}}_{B - L} \;\; , \;\;  Y^{\text{final}}_{L}= (c_s -1) Y^{\text{final}}_{B - L} \;\; ,
\end{align}    
where $c_s= 28 / 79$~\cite{Khlebnikov:1988sr, Harvey:1990qw} for non-supersymmetric models with one Higgs doublet
as assumed here.  

\begin{figure}[!htb!]
\centering
\includegraphics[width=0.32\textwidth]{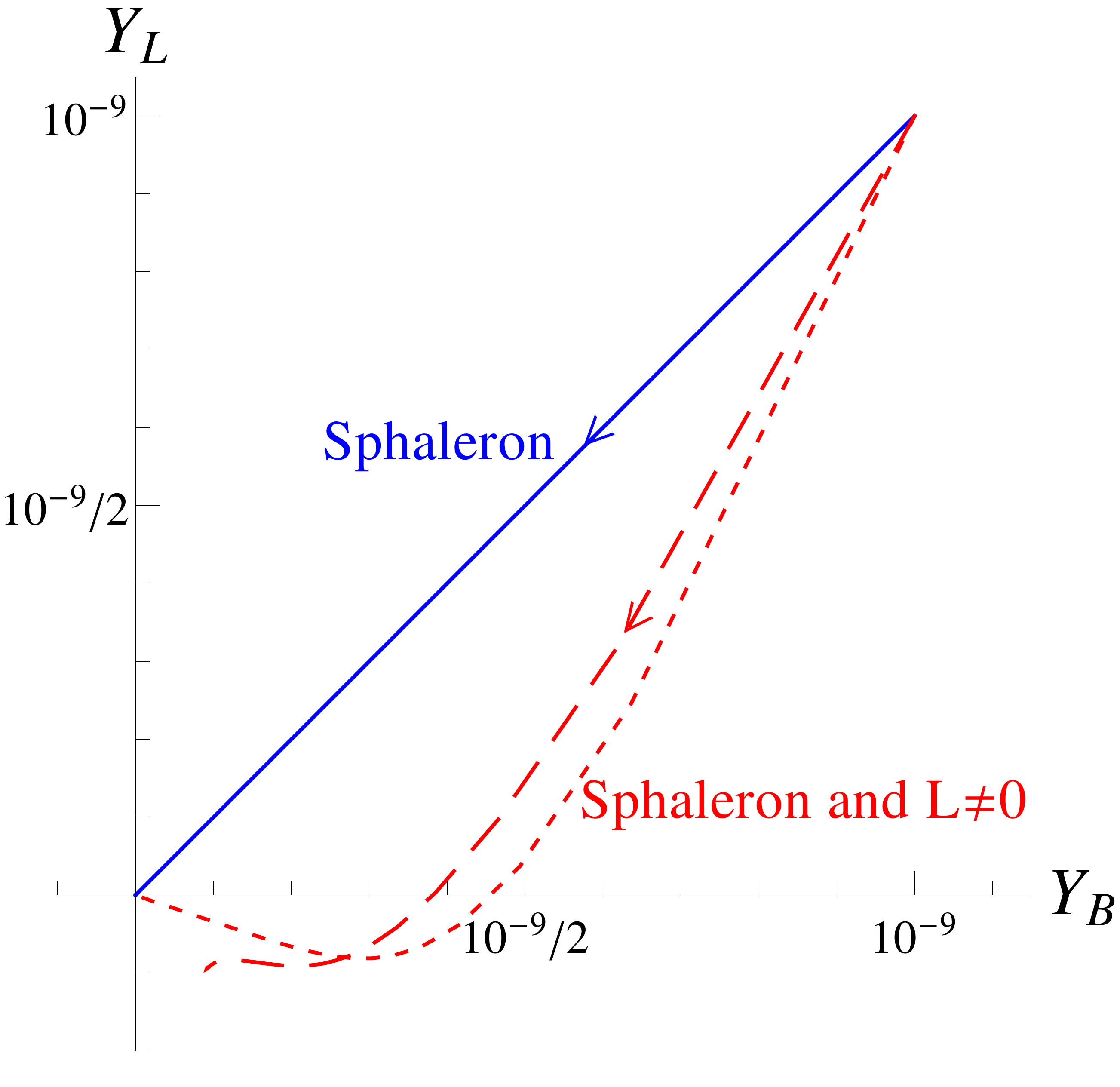}
\caption{The time evolution of an initial asymmetry of $Y_{B}=Y_{L}=10^{-9}$ with the sphaleron processes only (blue solid line),
and with the sphaleron processes plus the damped and undamped $\slashed{L}$ washout interactions (red dashed and doted lines, respectively). 
Only in the case of damped $\slashed{L}$ interactions can final nonzero $B$ and $L$ asymmetries be obtained.
} 
\label{fig:washout_BL}
\end{figure}

Before numerically solving the Boltzmann equations, we first study a few simplified examples to illustrate qualitatively the interplay between the washout and the sphaleron processes
as shown in Fig.~\ref{fig:washout_BL} that do ${\it not}$ use the real thermal rates $\gamma$ for $L$ washout interactions in Eqs.~(\ref{Y_B-L}-\ref{Y_H'}) above.
Assuming an initial asymmetry of $Y_{B}=Y_{L}=10^{-9}$, the solid blue line shows that the sphalerons wipe out completely
the initial asymmetry in the absence of any $\slashed{L}$ washout process. 
The red dashed and doted lines demonstrate the effects of damped and undamped $\slashed{L}$ washout processes, respectively.
In the damped case the $\slashed{L}$ interaction is assumed to produce a time-dependent lepton number violation
$dL/dt \sim - \exp \lee - m \cdot t \rii$, where $m$ is an arbitrary mass scale such that the product $m \cdot t$ is dimensionless. 
One can see that the $L$ asymmetry is eliminated faster than the $B$ asymmetry and thus consequently a net $B-L$ asymmetry is created. 
At later times, the damped $\slashed{L}$ interaction becomes
inefficient and the sphalerons yield $Y_B+Y_L=0$, leading to a non-vanishing baryon asymmetry: $Y_{B}=-Y_{L} \sim 10^{-10}$.
In the case of an undamped
$\slashed{L}$ interaction, the $\slashed{L}$ washout process in concurrence with  the $\cancel{B+L}$ sphalerons
eliminates both of the $L$ and $B$ asymmetries. This clearly illustrates that
in order to generate non-vanishing final $B$ and $L$ asymmetries the $\slashed{L}$ and $\cancel{B+L}$ interactions cannot coexist for too long.

\section{Numerical results}\label{sec:N_res}

\begin{figure}[!htb!]
\centering
\includegraphics[width=0.4\textwidth]{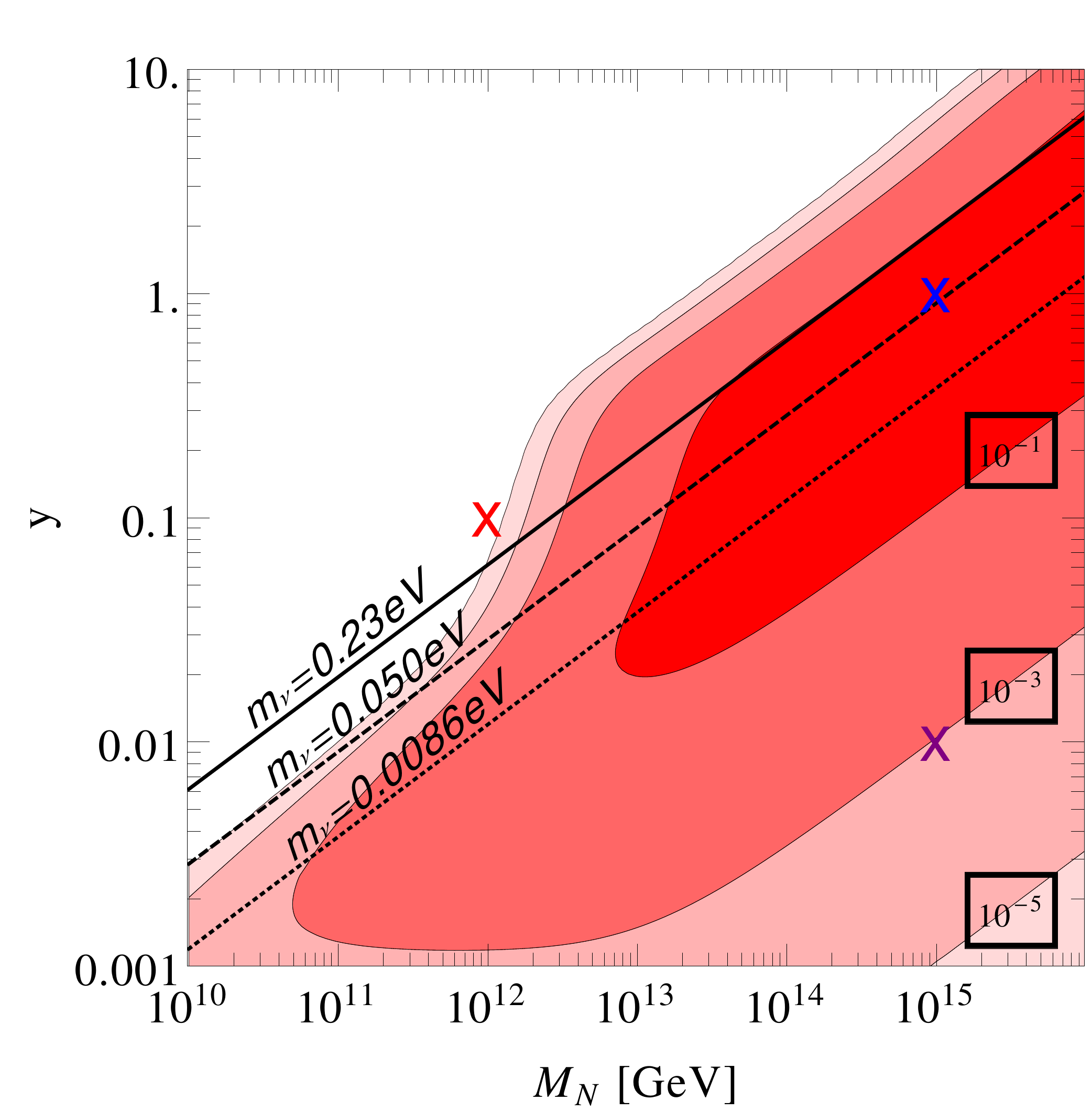}
\includegraphics[width=0.4\textwidth]{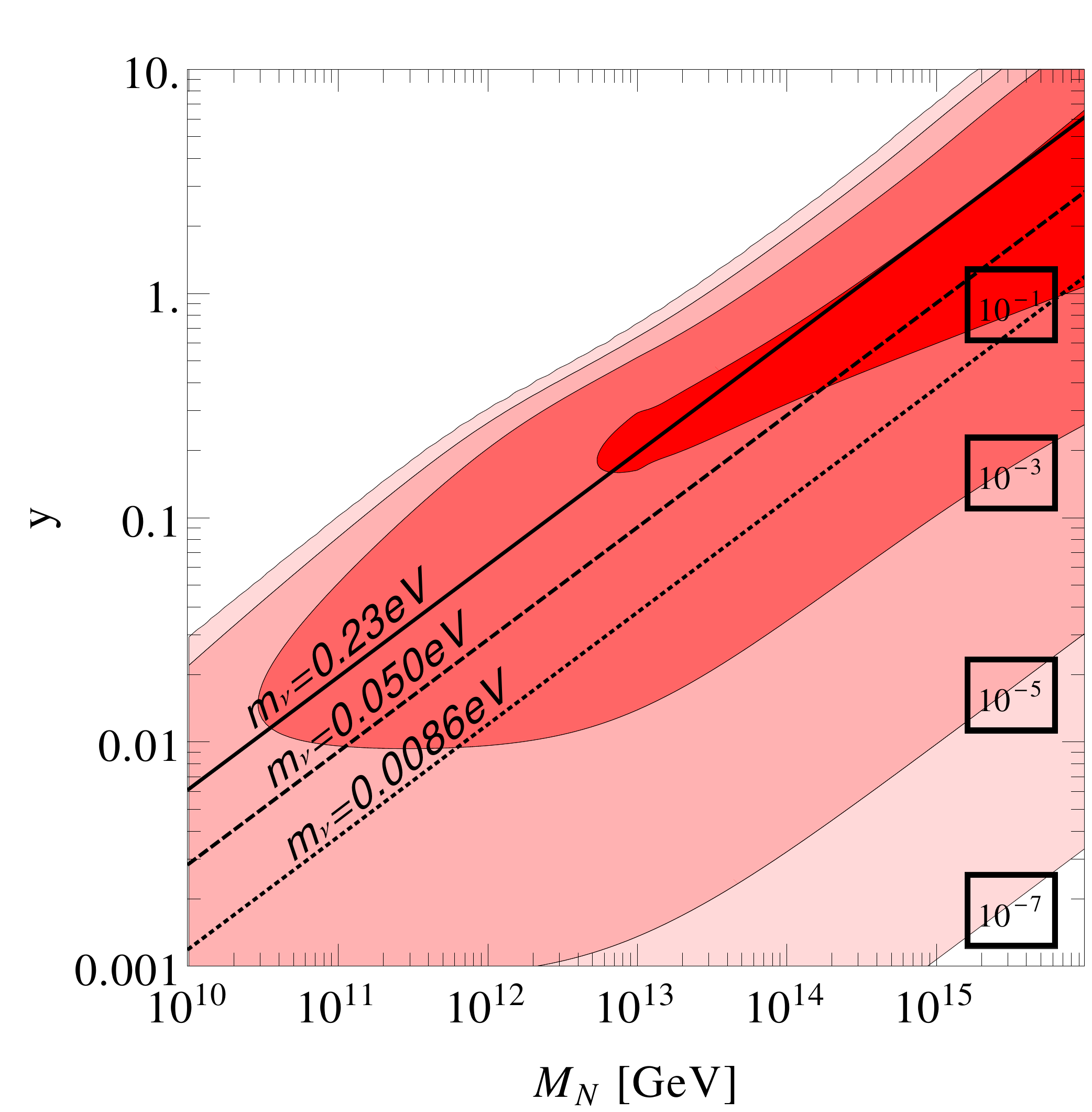}
\caption{Contour plots of $Y^{\text{final}}_{B-L}/Y^{\text{initial}}_{B+L}$ as a function of the Yukawa coupling $y$ (vertical axis) and the right-handed neutrino mass $M_N$ (horizontal axis). 
The initial $B+L$ generation is assumed to occur at $T=M_N/3$ (left panel) and $T=M_N/10$ (right panel).
Three diagonal straight lines, which correspond to different light neutrino masses resulting from the type-I seesaw mechanism, are shown.}
\label{fig:Boltz_BL}
\end{figure}

As mentioned before, the $\slashed{L}$ washout processes can convert an initial $B+L$ asymmetry originating from GUT baryogenesis
into a final $B-L$ asymmetry being immune against the action of the sphaleron processes. 
In Fig.~\ref{fig:Boltz_BL}, we present the numerical results of the ratio of the final $B-L$ to initial $B+L$ asymmetry, $Y^{\text{final}}_{B-L}/Y^{\text{initial}}_{B+L}$,
as a function of the Yukawa coupling $y$ (vertical axis) and the right-handed neutrino mass
$M_N$ (horizontal axis). The contours correspond to different values of $10^{-1}, 10^{-3}, 10^{-5}, 10^{-7}$ for the ratio $Y_{B-L}/Y_{B+L}$.

To evade complications from finite-temperature corrections, we confine ourselves to cases where the GUT $B+L$ injection scale is lower than $M_N$.
The conclusion, however, still holds when the scale is higher than $M_N$ for regions of maximal washout,
where $L$ washout is very efficient before the sphalerons become effective.
In the left panel the $B+L$ injection is assumed to take place at $T = M_N/3$ 
while the right panel assumes the injection scale to be $M_N/10$. 
The left and the right panels exhibit similar patterns but with two major differences. 
First, the right panel with the lower $B+L$ generation scale corresponds to a shorter period of $L$ 
washouts, that can be compensated by a larger Yukawa coupling~(and hence a higher $L$ washout rate).
As a result, the contours shift upward in the right panel compared to those in the left panel.   
Second, a lower injection scale corresponds to a smaller washout effect at the time of injection
since the resonance enhancement of $\slashed{L}$ processes occurs for $T \sim M_N$.
Depending on the sphaleron rate at the injection scale, the interplay between the  two interactions can have an impact on the final $B-L$ asymmetry. 
There exists, for instance, a noticeable difference in the bottom-left corner;
for the left panel, both $\slashed{L}$ and $\cancel{B+L}$ processes are strongly efficient at the time of $B+L$ injection,
leading to a smaller final $B-L$ asymmetry as indicated by the larger white area around the bottom-left corner as compared to the right panel.

\begin{figure}[!htb!]
\centering
\includegraphics[width=0.4\textwidth]{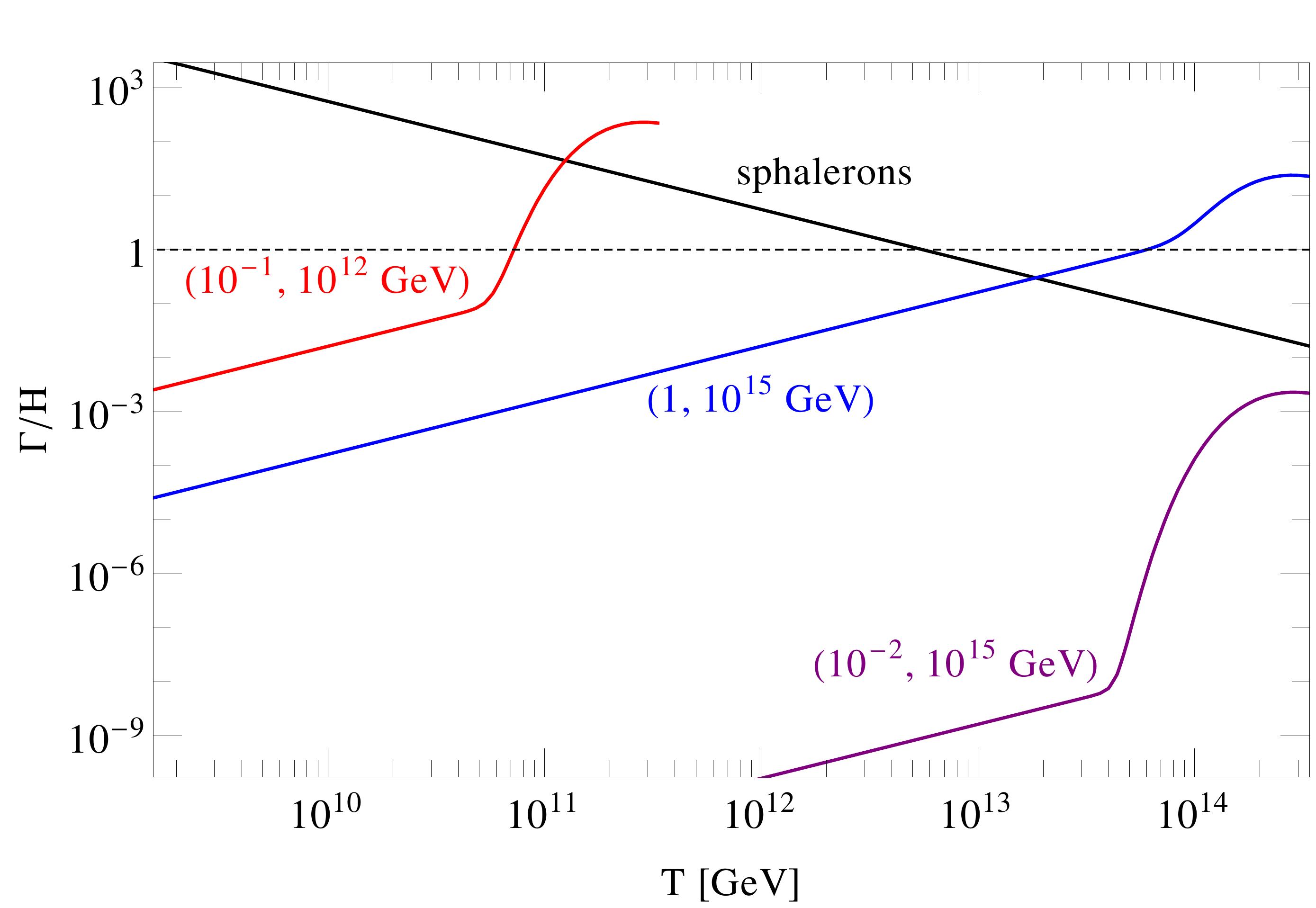}
\caption{The normalized sphaleron~(black straight line) and $L$ washout rates~(red/left, blue/middle, and purple/right curved lines) with respective to the Hubble expansion rate. 
The entries in the parentheses indicate $y$ and $M_N$, respectively. The three different washout rates correspond to the three crosses of the same color in the left panel of Fig.~\ref{fig:Boltz_BL}. }
\label{fig:H0_Ga}
\end{figure}

The final $B-L$ asymmetry depends strongly on the interplay and sequence of the washout and sphaleron processes.
The sequence has been illustrated in
Fig.~\ref{fig:H0_Ga}, where the interaction rates
of the washout and sphaleron processes, normalized to the Hubble expansion rate,
are displayed as functions of the temperature. The straight black line refers to the normalized sphaleron rate
while the curved lines correspond to the normalized $L$ washout rates in three benchmark cases,
marked by the colored crosses in Fig.~\ref{fig:Boltz_BL}.
The values in the parentheses refer to the Yukawa coupling $y$ and the right-handed neutrino mass $M_N$, respectively. 

A maximal $B-L$ asymmetry occurs around the top-right corner in Fig.~\ref{fig:Boltz_BL} where
the $\slashed{L}$ interactions are effective before the sphalerons become active, but they decouple when the sphaleron
processes come into equilibrium as shown by the blue/middle curved line in Fig.~\ref{fig:H0_Ga}. 
A small $B-L$ asymmetry results from two scenarios.
The first one corresponds to the bottom-right corner in Fig.~\ref{fig:Boltz_BL}, where the washout processes have never been fast enough
before the initial $(B+L)$ is eradicated by the sphalerons, as indicated by the right/purple curved line in Fig.~\ref{fig:H0_Ga}.  
The second one corresponds to the top-left region of Fig.~\ref{fig:Boltz_BL}
where the $L$- and $(B+L)$-violation processes coexist for a long time, as represented by the red/left curved line in Fig.~\ref{fig:H0_Ga}. 
In these two cases, both the $B$ and $L$ asymmetries are wiped out.

The common, general features for the normalized washout rates in Fig.~\ref{fig:H0_Ga} can be classified based on non-resonance~($T \ll M_N$)
and resonance regions~($T \sim M_N$). When the temperature is much lower than $M_N$, all the rates are linearly proportional to $T$:
\begin{align}
\frac{\Ga}{H} \sim \frac{y^4 T^3}{M^2_N} \frac{M_{Pl}}{T^2} = \frac{y^4 M_{Pl}}{M^2_N} T.
\end{align} 
The resonance enhancement appears when $T \sim M_N$ --
a smaller Yukawa coupling implies a narrower decay width of $N$, resulting in a larger enhancement.

From Fig.~\ref{fig:Boltz_BL}, for $ 10^{-2} \lesssim y  $ and $M_N \gtrsim 10^{13}$ GeV,
$Y^{\text{final}}_{B-L} / Y^{\text{initial}}_{B+L} \gtrsim 10^{-1}$ can be achieved.
For $M_N \lesssim 10^{12}$ GeV, $Y^{\text{final}}_{B-L} / Y^{\text{initial}}_{B+L}$ cannot exceed roughly $0.1$ regardless of values of $y$ as explained above.
Contingent upon the initial $B+L$ asymmetry, one can find regions of the parameter space to generate the desired $Y^{\text{final}}_{B-L} \simeq 2.4 \times 10^{-10}$ which corresponds to the observed baryon asymmetry, $Y^{\text{final}}_{B} = 8.7 \times 10^{-11}$~\cite{Ade:2015xua}.   
The black solid line represents the active neutrino mass of 0.23 eV that is the Planck bound on the sum of the active neutrino masses~\cite{Ade:2015xua},
while the dashed and dotted black lines denote
$m_\nu = \sqrt{\De m^2_{atm}} \simeq 0.05$ eV and $m_\nu = \sqrt{\De m^2_{sol}} \simeq 8.6 \times 10^{-3}$ eV~\cite{Agashe:2014kda}, respectively.
Therefore, to obtain the mass-squared difference corresponding to atmospheric neutrino oscillations from the type-I seesaw mechanism and
to achieve $Y^{\text{final}}_{B-L} / Y^{\text{initial}}_{B+L} \gtrsim 10^{-1}$
require a lower bound on
the right-handed neutrino mass: $M_N \gtrsim  10^{13}$ GeV with $ y\gtrsim 0.1$.
Note that the ratio $Y^{\text{final}}_{B-L} / Y^{\text{initial}}_{B+L}$ is independent of the value of $Y^{\text{initial}}_{B+L}$ since the Boltzmann equations are linear in $Y_B$ and $Y_L$.

In Ref.~\cite{Fukugita:2002hu} it was concluded that for $M_{N} \sim 10^{16}$ GeV and $y\sim 1$ the observed baryon asymmetry can be reproduced.
These values of $M_N$ and $y$ actually fall into the region of maximal washout, i.e., $Y^{\text{final}}_{B-L} / Y^{\text{initial}}_{B+L} \sim \mathcal{O}(0.1)$.
In other words, our numerical results of the $1+1$ scenario are consistent with those of Ref.~\cite{Fukugita:2002hu}.

\section{Generalization to a $3+2$ case}\label{sec:3+2}

Finally, we study a more general case including three generations of SM leptons and two right-handed neutrinos,
required to reproduce two mass-squared differences inferred from the solar and  atmospherical neutrino oscillations.

Before showing numerical results, we would like to comment on the properties of the $3+2$ case.
Here we mainly focus on the maximal washout region,
which as indicated above occurs for $M_N \gtrsim 10^{13}$ GeV.
As a consequence, none of the charged lepton Yukawa couplings is efficient in the regions of interest~($T \gtrsim 10^{13}$ GeV)
and so the three lepton generations are indistinguishable~\cite{Barbieri:1999ma,Davidson:2008bu}, implying that the neutrino mixing matrix~(PMNS matrix)
will not appear in the result.
Since one linear combination of $(\ell_e, \ell_\mu, \ell_\tau)$ will not couple to the right-handed neutrinos and hence remains massless, the lepton asymmetry stored along this direction will not be washed out.
In contrast, the two massive light neutrinos which couple to $N_{(1,2)}$ will participate in the washout processes and the asymmetries stored therein will be partially converted into the asymmetry of $H$. In other words, in the $3+2$ case only two linear combinations, denoted by $\ell_1$ and $\ell_2$, will give rise to a non-zero $(B/3-L_{(1,2)})$ and consequently also a final $B$, whereas the $(B/3 + L_3)$ asymmetry along the $\ell_3$ direction will be completely destroyed
by the sphalerons\footnote{The situation is quite similar to cases in leptogenesis where heavy neutrino flavor effects are important; not only $N_1$ but also $N_2$ contributes the $L$ asymmetry generation~\cite{Barbieri:1999ma,Nielsen:2001fy,Vives:2005ra,DiBari:2005st,Engelhard:2006yg}, depending on the flavor structure of Yukawa couplings $y^{\alpha j} \bar{\ell}_\alpha H N_i$~($\alpha=(e,\mu,\tau)$ and $i=(1,2)$) and on how strong the corresponding $L$ washout effects are.}.

In the following, we will not explore the full parameter space which is quite large
for the $3+2$ case. Instead a simplified framework, where each massive light neutrino couples
only to one of the heavy right-handed neutrinos but not to both of them, is chosen to highlight the main features.
The relevant Lagrangian is
\begin{align}
\mathcal{L} \supset y_1 \bar{\ell}_1 H^* N_1 + y_2 \bar{\ell}_2 H^* N_2 + \frac{1}{2} M_1 \overline{N^c_1}  N_1 +
 \frac{1}{2} M_2 \overline{N^c_2}  N_2 ,
\end{align}
where the neutrinos $\nu_1$ and $\nu_2$ in the $SU(2)_L$ doublets $\ell_1$ and $\ell_2$, respectively,
are the mass eigenstates of $m_1= y^2_1 v^2/M_1$ and
$m_2= y^2_2 v^2/M_2$ while $\nu_3$ in the doublet $\ell_3$ remains massless~($m_3 = 0$).
Due to the fact the PMNS matrix will not influence the washout calculation for $T> 10^{12}$ GeV, 
we only require $m_1$ and $m_2$ to reproduce the two mass-squared differences associated with solar and atmospheric neutrino oscillations.

With these assumptions the previous washout calculation is repeated, taking into account both the contributions of $N_1$ and $N_2$.   
For simplicity, we assume that the GUT $B+L$ injection scale is one third of the minimum of $M_1$ and $M_2$.
In addition, the normal mass hierarchy is presumed: $m_1=\sqrt{ \Delta m^2_{sol}} \simeq 8.6 \times 10^{-3}$ eV
and $m_2=\sqrt{\Delta m^2_{atm}} \simeq 0.05$ eV.

\begin{figure}[!htb!]
\centering
\includegraphics[width=0.4\textwidth]{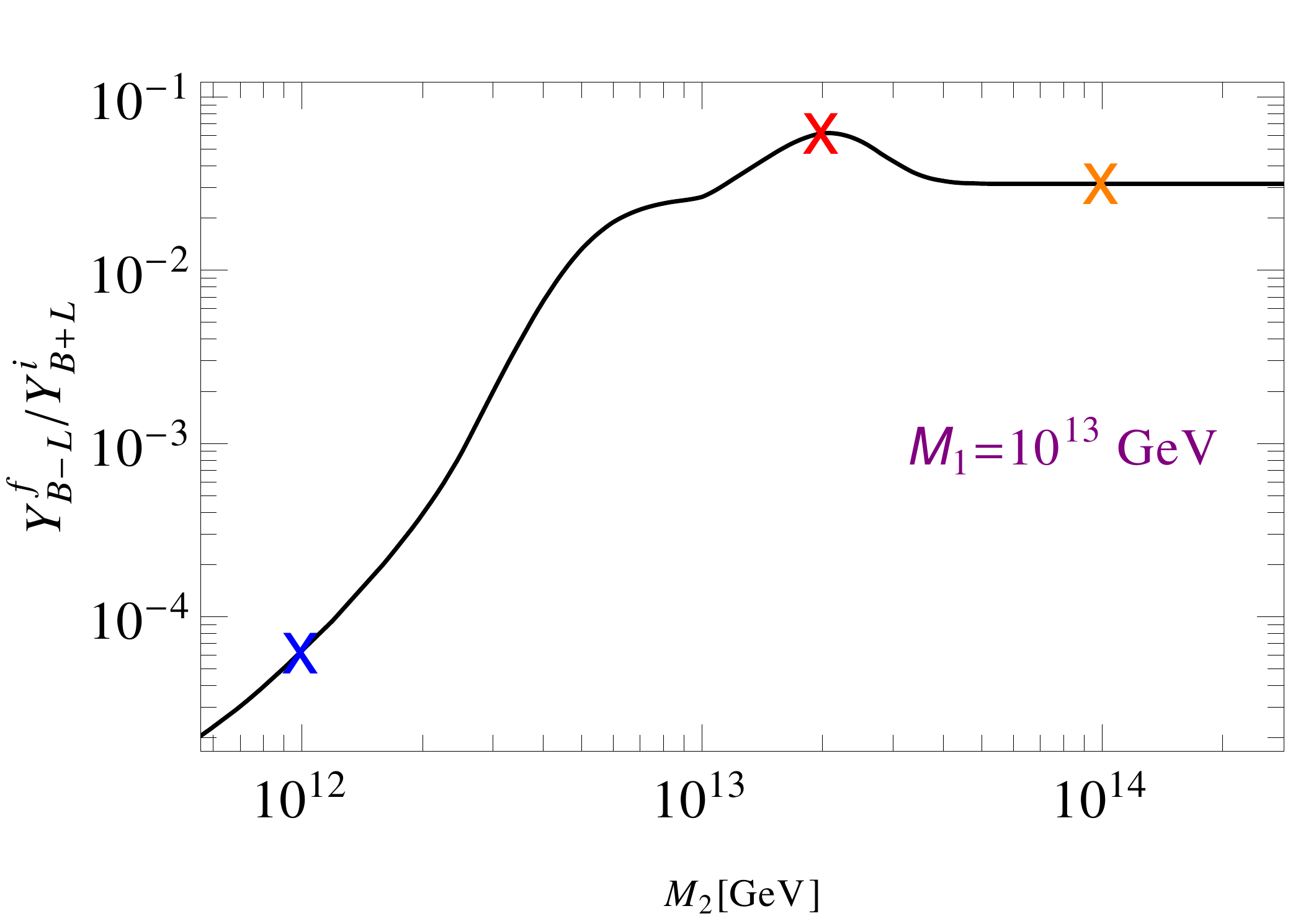}
\includegraphics[width=0.4\textwidth]{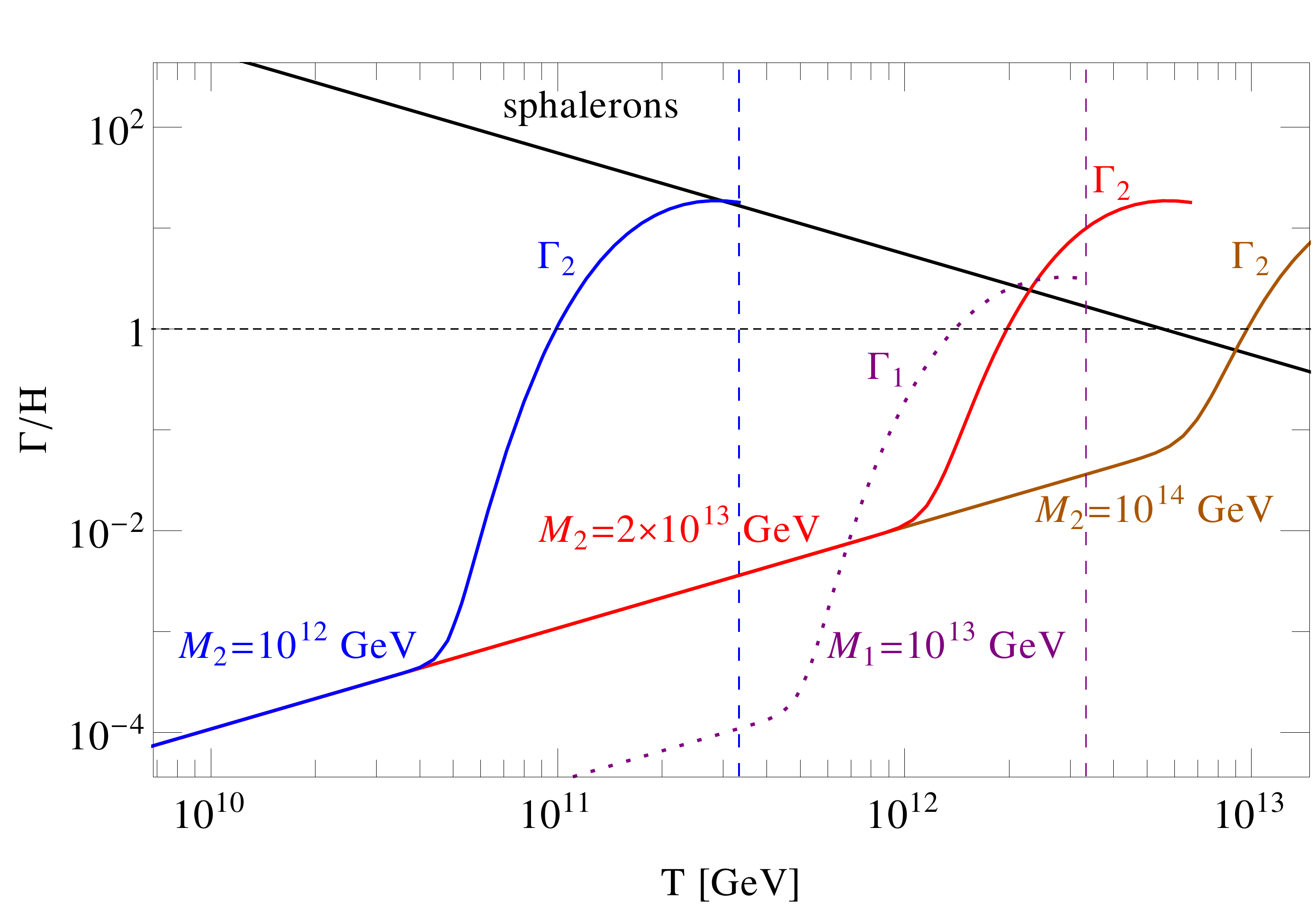}
\caption{Left panel: the ratio of final $B-L$ to initial $B+L$ versus $M_2$,
assuming $M_1=10^{13}$ GeV in the simplified $3+2$ case.
Right panel: similar to Fig.~\ref{fig:H0_Ga}, the normalized sphaleron and washout rates
for three representative points which are marked by the three crosses
in the left panel. The $N_2$-mediated washout rate $\Ga_2$ is indicated by the three solid lines,
corresponding to three different masses of $N_2$. Because of the fixed value of $M_1$, the $N_1$-induced washout rate $\Ga_1$
is denoted by the single purple dotted line. }
\label{fig:wash_32}
\end{figure}

Setting $M_1=10^{13}$ GeV, the ratio of $Y^{\text{final}}_{B-L}$ to $Y^{\text{initial}}_{B+L}$ is shown in the left panel of Fig.~\ref{fig:wash_32} 
as a function of $M_2$. The ratio is strongly suppressed for small $M_2$, peaks around $M_2 \sim 2 \times 10^{13}$ GeV with 
$Y^{\text{final}}_{B-L} / Y^{\text{initial}}_{B+L} \gtrsim 3 \times 10^{-2}$ and flattens for larger $M_2$.
Its behavior can be understood via the interplay between the sphaleron and washout interaction rates
as displayed in the right panel, where three cases corresponding to the three crosses in the left panel are presented.
The $N_2$-induced washout rate $\Ga_2$ is indicated by the solid lines
while the $N_1$-mediated one is denoted by the purple dotted line.
The vertical dashed lines mark the corresponding
$B+L$ injection scale which is chosen to be $\text{min}(M_1,M_2)/3$. For the two cases with $M_2 > M_1$,
the  scale is simply $M_1/3$.

For smaller values of  $M_2$, for example $M_2=10^{12}$ GeV indicated by the blue/left cross in the left panel of
Fig.~\ref{fig:wash_32}, both the sphaleron and $N_2$-mediated lepton number violation rates are effective as shown in
the right panel such that both the $B$ and $L_2$ asymmetries are erased.
On the other hand, the $\ell_1$-lepton washout rate, $\Gamma_1$, is never fast enough to create a sizable $B/3 - L_1$ asymmetry before the sphalerons wipe out an initial $B/3 + L_1$ asymmetry.
Consequently, neither $N_1$- nor $N_2$-mediated washout processes can generate a sizable $B-L$ asymmetry. 

For larger values of $M_2$ with $M_2 \sim M_1$, the $B+L$ injection scale becomes higher, 
implying a smaller sphaleron rate ($\Gamma_{\text{Sph}}/H \sim 1/T$) around the region of the 
washout resonance, $T \sim M_2$, as shown in the right panel. In this case, the substantial 
$B/3-L_{(1,2)}$ asymmetries are generated by effective $L_1$ and $L_2$ washouts while the sphaleron rate
is suppressed.

When $M_2$ becomes much bigger than $M_1$, the $B+L$ injection scale is fixed to be $M_1/3$.
In this region,  $L$ violation is created mainly by $N_1$ because of $\Ga_1 \gg \Ga_2$, leading to a sizable final $B/3 - L_1$ asymmetry but a tiny $B/3 - L_2$ asymmetry.
That is the reason why the maximal $Y^{\text{final}}_{B-L} / Y^{\text{initial}}_{B+L}$ becomes smaller than that of the previous case with $M_2 \sim M_1$
as seen by comparing the red/middle cross with the orange/right one in the left panel. 

To summarize, for regions with large $M_1$ and $M_2$ sizable $L$ washouts can be obtained and the observed
mass-squared differences can be realized in the simplified 3+2 case. 
For general $3+2$ cases, each light neutrino will receive both $N_1$ and $N_2$ 
contributions. The conclusions drawn above, however, remain valid as long as there is no severe cancellation
between the two contributions.


\section{Conclusions}~\label{sec:conclusion}
In this work, we have revisited the idea proposed in Ref.~\cite{Fukugita:2002hu}  
where lepton number violation, induced by the right-handed neutrinos, results in a nonzero $B-L$ asymmetry.
In other words, the lepton number violating processes can 
make part of an original baryon asymmetry created in GUT baryogenesis immune against the sphaleron interactions.
We have developed the scenario further by 
numerically solving the Boltzmann equations including both the $L$-violating washout and the $(B+L)$-violating sphaleron
processes.
The $L$ washout efficiency is defined by the ratio of the final $B-L$ to the initial $B+L$ asymmetry, $Y^{\text{final}}_{B-L} / Y^{\text{initial}}_{B+L}$.
Large final baryon asymmetries can be obtained
when the washout processes are in equilibrium
before the sphalerons become important but are ineffective when the sphalerons are at work. 
This corresponds to regions of large right-handed neutrino mass~($ M_N\gtrsim 10^{13}$ GeV). 

For the case of one active and one right-handed neutrino $N$, in the absence of quark Yukawa couplings a maximal $B-L$ asymmetry is
only one fourth of the initial $B+L$ asymmetry, which is due to chemical equilibrium between the lepton and Higgs doublets from the dominant $\Delta L=2$ processes.
Taking into account the top and bottom Yukawa couplings, which are
efficient for $T\lesssim 10^{16}$ GeV and $T \lesssim 10^{12}$ GeV respectively, the $L$ asymmetry can be further transferred into
quark asymmetries, thereby enhancing the $L$ washout effect: $Y^{\text{final}}_{B-L} / Y^{\text{initial}}_{B+L}=1/3$ ($t$ Yukawa coupling only)
and $Y^{\text{final}}_{B-L} / Y^{\text{initial}}_{B+L}=2/5$ ($t$ and $b$ Yukawa couplings).
Depending on the $B+L$ injection scale from the heavy Higgs or  gauge boson decays in the GUT baryogenesis,
$Y^{\text{final}}_{B-L} / Y^{\text{initial}}_{B+L}$ can be of $\mathcal{O}(0.1)$ for a right-handed neutrino  
mass $M_N$ above $10^{13}$ GeV and for Yukawa couplings $y \gtrsim 10^{-2}$.
The neutrino mass-squared differences corresponding to solar and atmospheric neutrino oscillations can be accommodated
within the region of maximal washout.

Moreover, we generalized the original $1+1$ scenario to a more realistic case of three active and two right-handed neutrinos, where one light neutrino
remains massless. In this case, there are contributions to washout processes from both heavy neutrinos $N_1$ and $N_2$.
Since we are interested in the region of maximal washout, $T\sim M_N  \gtrsim 10^{13}$ GeV, the three lepton generations are indistinguishable
and so the neutrino mixing matrix has no influence on the $L$-violating processes.
Nonetheless, the heavy neutrino flavor structure, i.e., how $N_1$ and $N_2$ couple to leptons plays a role. 
Instead of exploring the full parameter space, we have chosen a simplified framework where each light massive neutrino
couples to only one of the two right-handed neutrinos but not to both of them.
The corresponding two light neutrino masses are required to reproduce
the mass-squared differences related to solar and atmospheric neutrino oscillations. The conclusions are similar to those of the $1+1$ case.
For lower right-handed neutrino masses~($\lesssim 10^{12}$ GeV), the final $B-L$ asymmetry is small since the sphalerons and
$L$-violating processes are simultaneously active and eliminate both $B$ and $L$ asymmetries.
In regions with larger right-handed neutrino masses~($ \gtrsim 10^{13}$ GeV), the sphalerons have not come into equilibrium
while the $L$ washouts~(from either of $N_1$ and $N_2$ or both) are active.
Consequently, sizable conversion rates $\left( Y^{\text{final}}_{B-L} / Y^{\text{initial}}_{B+L} \gtrsim 10^{-1} \right)$ can be
achieved, leading to a substantial final $B-L$ and thus $B$ asymmetry.

\appendix
\section{Impacts of top and bottom Yukawa couplings}\label{app:Boltz} 

We here investigate how the existence of the heavy quark Yukawa couplings modify the Boltzmann equations.
In the following, it is always assumed that all asymmetries (quarks, leptons and Higgs bosons) are much smaller
than the equilibrium density of the corresponding particle. That is,
\begin{align}
Y_{\Delta f} = Y_{f} - Y_{\bar{f}} \ll Y^{eq}_f,
\end{align}
for $f= Q_3$~(third generation quark doublet), $U_3$~(right-handed $t$), $D_3$~(right-handed $b$), $\ell$ and $H$, where $Q_3$, $\ell$ and $H$ have two degrees of freedom from gauge multiplicity. 
To make the notation consistent with $Y_B$ and $Y_L$ which denote the baryon and lepton asymmetry densities,
we drop `$\Delta$' in $Y_{\Delta f}$, i.e., 
\begin{align}
Y_{\Delta f} \to Y_{f} \ .
\end{align}
In contrast, $Y^{eq}_f = Y^{eq}_{\bar{f}}$ always represents the equilibrium density of the (anti-)particle.

The Boltzmann equation for the lepton asymmetry, including the Yukawa interactions reads
\begin{align}
zHs \frac{d Y_{L}}{dz} = &2 \bigg[ - 2 \gamma_{Ns} \left( \frac{Y_{L}}{Y_{L}^{eq}} + \frac{Y_{H}}{Y_{H}^{eq}}\right) - 4\gamma_{Nt} \left( \frac{Y_{L}}{Y_{L}^{eq}} + \frac{Y_{H}}{Y_{H}^{eq}} \right) - \gamma_{Hs} \left( \frac{Y_{L}}{Y_{L}^{eq}} \frac{Y_N}{Y_N^{eq}} - \frac{Y_{Q_3}}{Y_{Q_3}^{eq}} + \frac{Y_{U_3}}{Y_{U_3}^{eq}} \right) \nonumber \\ &- \gamma_{Ht} \left( 2 \frac{Y_{L}}{Y_{L}^{eq}} + \left(\frac{Y_{U_3}}{Y_{U_3}^{eq}}+ \frac{Y_{Q_3}}{Y_{Q_3}^{eq}} \right) \left(1+\frac{Y_{N}}{Y_{N}^{eq}}\right) \right) -\gamma_{As} \left( \frac{Y_{L}}{Y_{L}^{eq}} \frac{Y_N}{Y_N^{eq}} + \frac{Y_{H}}{Y_{H}^{eq}}\right) \nonumber \\ &- \gamma_{At_1} \left( \frac{Y_{L}}{Y_{L}^{eq}} + \frac{Y_{H}}{Y_{H}^{eq}} \right) - \gamma_{At_2} \left( \frac{Y_{L}}{Y_{L}^{eq}} +\frac{Y_{H}}{Y_{H}^{eq}} \frac{Y_N}{Y_N^{eq}} \right) \bigg] \, ,
\label{eq:BolzDYL}
\end{align}
with $z=\frac{M_N}{T}$.

To simplify the calculation without including quarks in the Boltzmann equations,
one can assume that the relevant Yukawa couplings quickly transfer the asymmetry from $H$ to $Q_3$ and $U_3$ when $T \lesssim 10^{16}$ GeV,
and also to $D_3$ for $T \lesssim 10^{12}$ GeV.
In light of the chemical equilibrium of the Yukawa interactions and the correlations of the particle numbers among $H$, $Q_3$
and $U_3$~\footnote{Here we use the Maxwell-Boltzmann distribution which is a good approximation with the
average energy $\lan E \ran \sim 3 \ T$.}, it is straightforward to show that for $ 10^{12} \lesssim T \lesssim 10^{16}$ GeV
\begin{align}
\label{eq:t_YuKa}
Y_{H} = \frac{1}{2} Y_{H^{\prime}} \;\; , \;\; Y_{Q_3} = - \frac{1}{2} Y_{H^{ \prime } }  \;\; , \;\;  Y_{U_3} =  \frac{1}{2} Y_{H^{\prime}}
\;\; ,
\end{align}
where $Y_{H^{\prime}}$ is the total asymmetry obtained from the $L$ washout, i.e., $Y_{H^{\prime}} (z) \equiv
Y^{\text{initial}}_{L} - Y_{L}(z)$.
Similarly, for $T \lesssim 10^{12}$ GeV, one has
\begin{align}
\label{eq:b_YuKa}
Y_{H} = \frac{1}{4} Y_{H^{\prime}} \;\; , \;\; Y_{Q_3} = 0  \;\; , \;\;   Y_{U_3} =  \frac{3}{8} Y_{H^{\prime}}
\;\; , \;\;  Y_{D_3} =  - \frac{3}{8} Y_{H^{\prime}} \;\; .
\end{align}
From Eqs.~\eqref{eq:BolzDYL}, \eqref{eq:t_YuKa} and \eqref{eq:b_YuKa},
one recovers Eq.~\eqref{eq:dYL/dz} with
\begin{align}
b_{(H,~Q_3,~U_3)} = \left\lbrace \begin{matrix}
\frac{1}{2},-\frac{1}{2},~\frac{1}{2} &\;\;  \;\; 10^{12} \lesssim T  \lesssim 10^{16}\text{ GeV}\\
\frac{1}{4},~ 0,~\frac{3}{8}  &\;\; \;\; T \lesssim 10^{12}\text{ GeV}
\end{matrix} \right. \;\; .
\end{align}

Note that from Eqs.~\eqref{eq:t_YuKa} and \eqref{eq:b_YuKa}  if the $t$~($t$ and $b$) Yukawa coupling is in thermal equilibrium,
one has a maximal $Y^{\text{final}}_{B-L} / Y^{\text{initial}}_{B+L}$ of $1/3~(2/5)$ which is larger than $Y^{\text{final}}_{B-L} / Y^{\text{initial}}_{B+L}=1/4$, obtained by
ignoring the Yukawa couplings.

\section*{Acknowledgments}
We thank Evgeny Akhmedov for useful discussions.
This work is supported by DGF Grant No. PA 803/10-1.

\bibliography{GUT_N_revisit}
\bibliographystyle{h-physrev}

\end{document}